\documentclass[prx,twocolumn,superscriptaddress,floatfix]{revtex4-2}
\usepackage{bm,graphicx,url,epsf,color,amsmath,amssymb}
\usepackage{MnSymbol}
\usepackage{physics}

\begin{document}
\title{Heating Dynamics of Mesoscopic Electron Baths at High Magnetic Field}

\author{F.~Zanichelli}
\affiliation{Universit\'e Paris-Saclay, CNRS, Centre de Nanosciences et de Nanotechnologies, 91120 Palaiseau, France}
\author{A.~Veillon}
\affiliation{Universit\'e Paris-Saclay, CNRS, Centre de Nanosciences et de Nanotechnologies, 91120 Palaiseau, France}
\author{C.~Piquard}
\affiliation{Universit\'e Paris-Saclay, CNRS, Centre de Nanosciences et de Nanotechnologies, 91120 Palaiseau, France}
\author{A.~Aassime}
\affiliation{Universit\'e Paris-Saclay, CNRS, Centre de Nanosciences et de Nanotechnologies, 91120 Palaiseau, France}
\author{Y.~Sato}
\affiliation{Universit\'e Paris-Saclay, CNRS, Centre de Nanosciences et de Nanotechnologies, 91120 Palaiseau, France}
\author{A.~Cavanna}
\affiliation{Universit\'e Paris-Saclay, CNRS, Centre de Nanosciences et de Nanotechnologies, 91120 Palaiseau, France}
\author{Y.~Jin}
\affiliation{Universit\'e Paris-Saclay, CNRS, Centre de Nanosciences et de Nanotechnologies, 91120 Palaiseau, France}
\author{J.~Folk}
\affiliation{Department of Physics and Astronomy, University of British Columbia, Vancouver, British Columbia, V6T1Z1, Canada}
\author{U.~Gennser}
\affiliation{Universit\'e Paris-Saclay, CNRS, Centre de Nanosciences et de Nanotechnologies, 91120 Palaiseau, France}
\author{A.~Anthore}
\email[e-mail: ]{anne.anthore@c2n.upsaclay.fr}
\affiliation{Universit\'e Paris-Saclay, CNRS, Centre de Nanosciences et de Nanotechnologies, 91120 Palaiseau, France}
\affiliation{Universit\'{e} Paris Cit\'{e}, CNRS, Centre de Nanosciences et de Nanotechnologies, F-91120 Palaiseau, France}
\author{F.~Pierre}
\email[e-mail: ]{frederic.pierre@cnrs.fr}
\affiliation{Universit\'e Paris-Saclay, CNRS, Centre de Nanosciences et de Nanotechnologies, 91120 Palaiseau, France}

\begin{abstract}
Quantum thermodynamics addresses the dynamics of heat flow in quantum devices driven out of equilibrium. 
Although mesoscopic circuits at low temperatures provide a flexible platform to explore this dynamics, experimental studies are wanting because thermal timescales in nanodevices are often too fast.
Here we engineer and investigate with noise thermometry a mesoscopic thermal circuit where heat flows between electron, phonon and nuclear systems can occur on slower timescales.
The central constituent of this device is a micrometer-scale metallic island electrically connected to large cold electron reservoirs through two to four ballistic quantum Hall channels, a component frequently used for exploring stationary thermal currents.
We uncover a two-step thermalization process specific to the mesoscopic scale, involving a fast initial temperature step followed by a much slower rise extending over minutes.
This observation is quantitatively accounted for by the balance between heat flows through electronic quantum channels, to cold phonons, and to the nuclear spins in the metallic island. The disclosed mesoscopic thermalization takes a step into the field of quantum thermo-\emph{dynamical} phenomena, highlighting their distinctive nature on a central constituent of quantum circuits. 
The implications for the thermal engineering of nanodevices include the thermal characterization of exotic states at high magnetic field.
\end{abstract}

\maketitle

\section{INTRODUCTION}

Mesoscopic circuits at low temperatures constitute a particularly interesting platform to experimentally investigate quantum thermodynamics \cite{Pekola2015}.
With intermediate dimensions between the microscopic and macroscopic scales (typically in the nanometer and micrometer ranges), quantum effects such as coherence, superposition, and entanglement can be important.
Pioneering experimental works have probed various thermodynamic properties of quantum circuits, such as thermopower in quantum point contacts \cite{Molenkamp1990,Molenkamp1992} and quantum dots \cite{Staring1993,Dzurak1993} (for a theoretical review, see e.g.\ \cite{Sothmann2015,Benenti2017}), as well as quantum heat transport \cite{Schwab2000,Meschke2006,Chiatti2006,Jezouin2013b,Majidi2024} (for a review, see \cite{Pekola2021}).
However, these studies remain mostly restricted to the stationary regime (see e.g.\ \cite{Gasparinetti2015,Wang2019} for rare exceptions), hence excluding dynamical processes at the heart of quantum thermodynamics, as well as excluding the temporal fluctuations that emerge at small system sizes.

\begin{figure}
\centering\includegraphics[width=1\columnwidth]{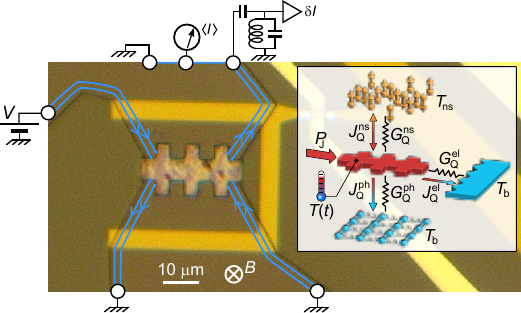}
\caption{
\footnotesize
Optical microscope image of a measured device.
A micron-scale metallic island is electrically connected via edge states (blue lines with arrows) of a two-dimensional electron gas (2DEG, slightly darker areas) in the integer quantum Hall regime  ($B\simeq4.8$\,T at filling factor $\nu=2$, $B\simeq10.8$\,T at filling factor $\nu=1$).
In the displayed configuration of $\nu=2$ and unbiased capacitively coupled top gates (orange), there are in total $N=4$ connected edge channels (one incoming line and one outgoing line per channel).
A Joule power $P_\mathrm{J}(V)$ is dissipated into the island's electron bath and the resulting temperature rise $T-T_\mathrm{b}$ is obtained from the increase in current fluctuations $\Delta S_\mathrm{I}\equiv\langle \delta I^2 \rangle (V)-\langle \delta I^2 \rangle (0)$.
Inset: thermal schematic. 
The island’s electron bath (red) is coupled with the phonon bath at base temperature $T_\mathrm{b}$ (blue, bottom), with external electronic baths also at $T_\mathrm{b}$ (blue, right) through the edge channels, and with the nuclear spin bath at temperature $T_\mathrm{ns}$ (orange).
\normalsize
}
\label{fig-sample}
\end{figure}

Here we investigate the heating dynamics of the electron bath formed within a micrometer-size metallic island, electrically connected to cold electron reservoirs through a few ballistic quantum channels.
In this mesoscopic regime, balanced contributions of the heat flows of electronic quantum channels ($J_\mathrm{Q}^\mathrm{el}$) and of electrons to nuclear spins ($J_\mathrm{Q}^\mathrm{ns}$) can give rise to a novel, two-step thermal dynamics. 
Such behavior may appear counterintuitive given the typically weak hyperfine interaction between electrons and nuclear spins, which is known to allow for a relatively large quantum coherence of spin qubits in mesoscopic systems \cite{Burkard2023}. 
Moreover, this observation  contrasts with previous investigations of stationary heat transport in comparable mesoscopic islands, where no significant contribution from $J_\mathrm{Q}^\mathrm{ns}$ was reported \cite{Jezouin2013,Sivre2018,Banerjee2017,Banerjee2018}.
However, the growing heat capacity of nuclear spins with the magnetic field is known to dominate the thermal behavior of macroscopic metallic systems, with applications such as adiabatic nuclear demagnetization cooling \cite{Pobell2007}.
In the presently explored mesoscopic scale, the nuclear spins are neither negligible nor completely dominant, and specifically manifest themselves in the thermal dynamics.

The configuration implemented in this work is very similar to one previously used to probe the steady-state electronic heat current for different (integer and fractional) quantum Hall filling factors \cite{Jezouin2013b,Banerjee2017,Sivre2018,Srivastav2021,LeBreton2022}, but here it is employed to address the dynamical regime.
As illustrated schematically in Fig.~\ref{fig-sample}, the electrons in the island are coupled to phonons through the deformation potential, to electrons in large contacts through the connected electronic channels and to the local nuclear spins through the hyperfine interaction.
The electron, phonon and nuclear spin baths can coexist in the island with different but well-defined temperatures.
Indeed, in metals at sub-Kelvin temperatures, electron-electron, phonon-phonon and nuclear spin-nuclear spin thermalization times range from nanoseconds to milliseconds, which are significantly shorter than the corresponding thermalization times between these different baths \cite{DelFatti2000, Pierre2003, Pobell2007, Turrell1988}.
The heating dynamics of the electron bath is obtained by monitoring the evolution of the electronic temperature in the island after a quick change of the applied Joule power. 
Specifically, the electrons in the island are heated up (or cooled down) with a well-defined step up (or down) in Joule power.
The resulting increase (decrease) in their temperature $T$ is then directly measured by noise thermometry.
With this approach, we find that the electron thermalization within the mesoscopic island occurs in two steps: an initial fast jump in temperature within the first second, followed by a slower change that extends over several minutes. 
The presence of these two stages and their relative temperature amplitudes is found to result from the competition between outgoing heat flows (through electronic quantum channels and toward phonons) and heat redistribution between electrons and nuclear spins within the island.
Because nuclear spins are coupled primarily to electrons 
(their interaction with phonons is very small \cite{Abragam1961,Pobell2007}, which is also attested by the absence of a discernible effect on stationary heat currents), their role remained hidden in previous steady-state measurements where nuclear spins and electrons in the island have reached the same temperature.
This is not the case in the dynamical regime where their thermal contribution is here evidenced.

\section{MEASURING AN ELECTRON BATH HEATING DYNAMICS}

The sample is nanofabricated on a Ga(Al)As heterojunction hosting a two-dimensional electron gas (2DEG) buried $90$\,nm below the surface. 
It is mounted in a dilution refrigerator with filtered and thermalized measurement lines \cite{Sivre2018}, cooled down to an electronic base temperature $T_\mathrm{b}$ in the 10\,mK range measured by on-chip noise thermometry (see Appendix A).
Two metallic islands of volume $\Omega=85\,\mu\mathrm{m}^3$ (see Fig.\ \ref{fig-sample}) and $3.7\,\mu\mathrm{m}^3$ electrically connected with the 2DEG are realized from a thermally annealed AuGeNi alloy (see Appendix A for precise composition).
The sample is immersed in a strong perpendicular magnetic field $B\simeq4.8$\,T or $10.8$\,T corresponding to the integer quantum Hall effect at filling factor $\nu=2$ or $\nu=1$.
In this regime, the current propagates along quantum Hall edge channels schematically shown as blue lines with arrows indicating the propagation directions.
The connection between these edge channels and the electron bath is effectively perfect, even for the smallest island (reflection probability $<0.3\%$, Appendix~A).
Gates capacitively coupled to the 2DEG underneath (orange in Fig.~\ref{fig-sample}) allow us to control by field effect the number $N$ of electronic channels connected to the island.
Two configurations were experimentally accessible at $\nu=2$: Either two edge channels (as illustrated in Fig.~\ref{fig-sample}) or one channel were transmitted under the orange gates, within each of the two 2DEG mesa separately connected to the island.
These mesa appear above and below the island in Fig.~\ref{fig-sample}; areas where the mesa is etched away appear slightly lighter.
In total, $N=4$ quantum Hall edge channels are connected in the first configuration; in the second $N=2$ channels are connected, with the inner quantum Hall edge channel reflected back to the island.

When applying a dc bias voltage, $V$, to the contact on the left side in Fig.~\ref{fig-sample}, the Joule power dissipated into the electron bath of the island is $P_\mathrm{J}=\tfrac{V^2}{2}\tfrac{Ne^2}{4h}$ with $e$ the elementary charge and $h$ the Planck constant \cite{Jezouin2013b}.
The resulting electron temperature increase $T-T_\mathrm{b}$ is determined experimentally from the increase in the measured spectral density of the current noise  $\Delta S_\mathrm{I}\equiv\langle \delta I^2 \rangle (V)-\langle \delta I^2 \rangle (0)$. 
For the noise signal propagating along $N/2$ outgoing edge channels, the temperature change is straightforwardly obtained from the well-established relation \cite{Buttiker1990,Jezouin2013b}
\begin{equation}
    \Delta S_\mathrm{I}=2 k_\mathrm{B}(T-T_\mathrm{b})Ne^2/4h,
    \label{Eq-NoiseToT}
\end{equation}
where $k_\mathrm{B}$ is the Boltzmann constant, and $Ne^2/4h$ corresponds to the conductance across the island.

\section{FIRST EVIDENCES FOR A TRANSIENT HEAT FLOW}

The temporal response of the island's electron bath temperature is given by
\begin{equation}
C_\mathrm{e} \frac{d T}{d t}=P_\mathrm{J}-J_\mathrm{Q},
\label{Eq-ElectronDynamics}
\end{equation}
where $C_\mathrm{e}$ is the heat capacity of the electronic bath, $P_\mathrm{J}$ the injected Joule power, and $J_\mathrm{Q}$ the total instantaneous outgoing heat current to cold electronic and phonon reservoirs, and toward the island's nuclear spins. 
There are two main differences between the dynamical and stationary regimes.
First, the temperature dynamics directly involves the electronic heat capacity, as shown in Eq.~\eqref{Eq-ElectronDynamics}, whereas the left-hand side cancels in the steady-state so incoming and outgoing powers must balance.
Second, it is important to consider transients in the heat flow, for example the contribution of nuclear spins to $J_\mathrm{Q}$ is nonzero only as long as they are not at equilibrium with the island's electrons.
Note that the chirality of edge currents protects both $J_\mathrm{Q}$ in the island's thermal balance Eq.~\eqref{Eq-ElectronDynamics} as well as the extraction of $T$ from Eq.~\eqref{Eq-NoiseToT} against downstream effects, such as electron cooling over long edge paths \cite{Lesueur2010,Bocquillon2013,Rosenblatt2020} (see Appendix~E for a confirmation from stationary heat currents and Appendix~G for further discussions).

We first consider thermal dynamics in the exclusive presence of the electronic and electron-phonon thermal couplings previously established in the stationary regime, estimating an upper limit for the corresponding thermalization time $\tau$ associated with the heat-up of the electron system.
It was shown in \cite{Jezouin2013b} that the heat current $J_\mathrm{Q}$ reduces in the stationary regime to the sum $J^\mathrm{el}_\mathrm{Q}(T,T_\mathrm{b})+J^\mathrm{ph}_\mathrm{Q}(T,T_\mathrm{b})$.
The first term, $J^\mathrm{el}_\mathrm{Q}$, is the heat flow from the electron bath at temperature $T$ toward cold electron reservoirs at base temperature $T_\mathrm{b}$, transmitted via the $N$ connected electronic channels.   
The second term, $J^\mathrm{ph}_\mathrm{Q}$, represents the heat flow from electrons in the islands to the cold phonon bath at $T_\mathrm{b}$.
To set an upper bound on $\tau$, we consider the case of only one connected electronic channel and neglect the parallel heat flow to phonons.
The resulting lower bound to the heat current $J_\mathrm{Q}$ then reduces to the quantum limit of the electronic heat flow along one channel, $J_\mathrm{K}=\frac{\pi^2 k_\mathrm{B}^2}{6h}(T^2-T_\mathrm{b}^2)$.
In the linear regime, where $T-T_\mathrm{b} \ll T$, this becomes $J_\mathrm{K}\simeq G_\mathrm{K}(T-T_\mathrm{b})$, with $G_\mathrm{K}=\frac{\pi^2 k_\mathrm{B}^2 T}{3h}$ the thermal quantum of conductance.
The characteristic timescale for the temperature evolution is then given by $\tau=C_\mathrm{e}/G_K$.
For electrons at temperature $T$ in a metal, the electronic heat capacity is $C_\mathrm{e}=(\pi^2 \nu_\mathrm{F} \Omega k_\mathrm{B}^2 T)/3$, where $\Omega$ is the volume of the reservoir and $\nu_\mathrm{F}$ is the density of states per unit of volume around the Fermi level.
Assuming a typical value of $\nu_\mathrm{F}\sim10^{47}\,\mathrm{J}^{-1}\mathrm{m}^{-3}$, as found in good metals such as gold (the main constituent of the metallic islands), we find $\tau/\Omega=h\nu_\mathrm{F} \sim100\,\mu\mathrm{s}/\mu\mathrm{m}^3$.  
In our largest island, this estimate yields a thermal relaxation time of at most 10\,ms--a value that is much shorter than the 1\,s experimental time resolution of our setup.
This hierarchy holds up to electronic density of states higher by two orders of magnitude with respect to gold, encompassing most materials.

\begin{figure}[htb]
\centering\includegraphics[width=1\columnwidth]{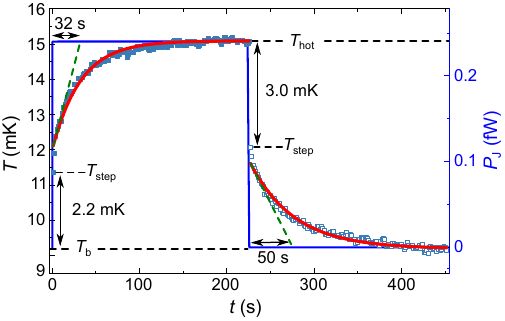}
\caption{
\footnotesize
Electronic temperature evolution in a mesoscopic island submitted to a change of Joule heating.
The metallic island (volume $\Omega\simeq85\,\mu\mathrm{m}^3$) is connected via $N=4$ electronic channels to cold reservoirs at $T_\mathrm{b}=9.2$\,mK. 
It is heated with $P_\mathrm{J}\simeq0.24$\,fW for $t\in[0,225]$\,s before cooling back to $T_\mathrm{b}$ with $P_\mathrm{J}=0$ at $t>225$\,s (see blue line and right vertical axis for applied $P_\mathrm{J}$).
The experimental electronic temperature $T(t)$ (symbols) is derived from noise measurements over $1$\,s interval using Eq.~\eqref{Eq-NoiseToT}.
Symbol sizes represent the standard error from averaging $430$ individual traces.
Red lines are exponential fits with the characteristic time $\tau$ and the height of the slow temperature evolution as parameters (see Appendix B).
The green dashed lines show the asymptotic slopes of the exponential functions at short times, crossing the steady-state temperature after a time $\tau$.
\normalsize
}
\label{fig-ResponseTime}
\end{figure}

Figure~\ref{fig-ResponseTime} shows a representative observation of the time evolution of the electron temperature $T$ measured in the 85\,$\mu\mathrm{m}^3$ metallic island ($0.34\,\mu$m of AuGeNi deposited over an area of $250\,\mu\mathrm{m}^2$).
In contrast to the estimate above, the measured temperature clearly evolves on timescales much longer than the experimental time resolution.
The continuous blue line in Fig.~\ref{fig-ResponseTime} shows the time-dependent power applied to the system: Starting from base temperature $T_\mathrm{b}\simeq9.2$\,mK, a quick Joule power jump ($\sim10$\,ms) from $0$ to $0.24$\,fW occurs at $t=0$, with a quick jump back to 0\,fW at $t=225$\,s.
The data (symbols) illustrate thermal equilibration in two steps, with very different timescales.
First, $T$ steps up instantly (compared to the 1\,s integration window of thermal noise) by $2.2$\,mK concomitant with the application of $P_\mathrm{J}=0.24$\,fW.
Second, $T$ increases slowly by an additional $3.7$\,mK, extending over half a minute to reach the steady-state (`final') temperature $T_\mathrm{final}\simeq15.1$\,mK.
A similar two-step behavior of $T(t)$ is observed when $P_\mathrm{J}$ is set back to 0, although with a noticeably longer time to reach the lower steady-state (base) temperature.
Exponential fits to the slower segments (red lines) give characteristic thermal relaxation times of $\tau\simeq32$\,s for warming up to $15.1$\,mK, and $\tau\simeq50$\,s for cooling down to $9.2$\,mK. 
As shown in the next section, the difference between the heating and cooling timescales comes from the fact that $\tau$ mainly depends on the final temperature.
Whereas the first fast $T$ step is compatible with the 10\,ms timescale expected from the electron-electron and electron-phonon mechanisms relevant in the steady-state, the second slow $T$ increase is not, by more than 3 orders of magnitude.
This indicates the presence of an additional heat transfer mechanism, beyond those evidenced in the stationary regime.
As demonstrated in the following, the second slow step is associated with the thermalization between electrons and nuclear spins in the island.
In the mesoscopic regime that is relevant here, with micrometer length scales and only a few quantum channels of electronic heat flow, both the relative amplitude of the slow temperature step as well as the long timescale itself can depend strongly on experimental parameters.

\section{THERMALIZATION TIME}

\begin{figure}[hb]

\centering\includegraphics[width=1.\columnwidth]{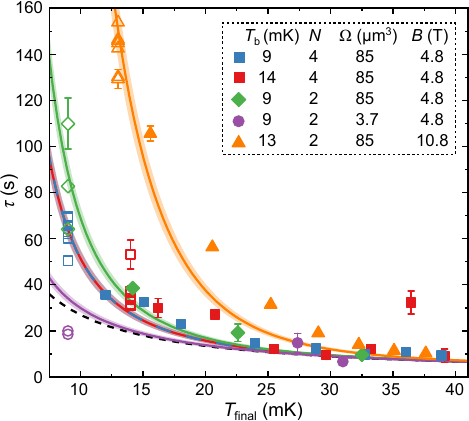}
\caption{
\footnotesize
Thermalization times $\tau$ versus experimental parameters.
Symbols: $\tau$ data points plotted as a function of the steady-state temperature $T_\mathrm{final}$. Full/open symbols correspond to a temperature step up/down (increasing/decreasing toward $T_\mathrm{final}$). 
Different symbol shapes indicate variations in the number of channels $N$, island volume $\Omega$, or magnetic field $B$.
Multiple identical open symbols at nearby $T_\mathrm{final}$ correspond to cooling from different hot temperatures.
Error bars display the standard error on $\tau$ when larger than symbol size.
Black dashed line: expected $\tau=K/T_\mathrm{final}$ in the limit of a dominant nuclear spin heat contribution and small magnetic energy $\mu B/k_\mathrm{B}T_\mathrm{final}\ll1$, with the Korringa coefficient $K=0.27$\,s.K.
Colored lines: predictions of Eq.~\eqref{Eq-TauGeneral}.
Semitransparent surroundings display the standard uncertainty on predictions from the separately characterized stationary heat conductance $G_\mathrm{Q}^\mathrm{stat}$, nuclear spin ratio $\gamma_\mathrm{ns}/K$ and $T_\mathrm{b}$.
}
\label{fig-TauVersusNandV}
\end{figure}

\subsection{Observations and underlying mechanism}

We now investigate the underlying mechanism responsible for the slow thermal relaxation. 
To identify it unambiguously, we extract the characteristic relaxation time while varying, \textit{in situ}, several discriminating parameters--$N$, $T$, $\Omega$ and $B$--that selectively impact different processes (see Appendixes~C and D).
For pure electron or phonon thermalization, one expects $\tau_\mathrm{el(ph)} = C_\mathrm{e}/G_\mathrm{el(ph)}$, where $C_\mathrm{e}$ is proportional to $\Omega$ and $T$, and $G_\mathrm{el(ph)}$ is the electron (phonon) thermal conductance. 
The electronic thermal conductance of $N$ parallel channels is also proportional to the temperature, and to $N$ (neglecting heat Coulomb blockade).
As a result, $\tau_\mathrm{el}$ decreases with increasing $N$, scales with the island volume $\Omega$, and is independent of both $T$ and the magnetic field $B$.
In contrast, for electron-phonon heat transfer, $G_\mathrm{Q}^\mathrm{ph}$ scales with temperature at a power close to $4$, and with the volume.
Therefore, $\tau_\mathrm{ph}$ decreases strongly with increasing $T$, but is essentially independent of $N$, $B$ and $\Omega$.
Only the electron-nuclear spin thermalization is affected by $B$, since both the nuclear spin heat capacity and the thermal conductance between nuclear spins and electrons depend on $B$.
The thermalization times $\tau$, extracted from exponential fits to the $T(t)$ data at different values of $N$, $T$, $\Omega$ and $B$, are plotted in Fig.~\ref{fig-TauVersusNandV} as a function of $T_\mathrm{final}$ (symbols).

A careful consideration of these data leads to three observations:
\textit{(i)} $\tau$ decreases rapidly with $T_\mathrm{final}$ at the smallest temperature and then much more slowly at higher temperatures $T_\mathrm{final}\gtrsim30$\,mK.
A decrease of $\tau$ with increasing $T$ is generally expected for electron-phonon heat transfers, but the rather weak temperature dependence above $T_\mathrm{final}\approx30$\,mK is not compatible with this strongly $T$-dependent mechanism.
\textit{(ii)} Changing only $N$, the number of ballistic channels, does not significantly impact $\tau$.
In contrast, if the electronic heat flow determined $\tau$, we would have expected a marked change, of at least 50\%.
We can also see that a strong reduction of the island's volume, by a factor of $25$, does not have a similar impact on the thermal relaxation time. 
This is once more in marked contrast with a slow heat transfer dominated by electronic heat current, for which $\tau$ would scale with $\Omega$.
\textit{(iii)} Raising the magnetic field by a factor of $2.2$ leaves $\tau$ mostly unchanged at $T\gtrsim20$\,mK but increases it significantly at lower $T$.

These observations provide a beam of evidence consistently pointing to an underlying electron-nuclear spin heat redistribution as the main mechanism responsible for determining the long relaxation time (see Appendix~G for further discussion, and Appendix~H for additional data on a different sample over an extended range of magnetic field).
At an approximate level, for a small enough nuclear magnetic moment $\mu\ll k_\mathrm{B}T/B$, the electron-nuclear spin relaxation time in metals is given by $\tau\simeq K/T$, with the Korringa coefficient $K\sim1$\,$\mathrm{s.K}$ (see e.g.\ \cite{Pobell2007,Abragam1961}).
At 10\,mK, this gives $\tau\sim100$\,s, in order of magnitude agreement with observations.

\subsection{Quantitative comparison}

In the following section, we aim for a quantitative understanding of this phenomenon.
The temperature evolution of the nuclear spin bath is described by:
\begin{equation}
    C_\mathrm{ns}\frac{d T_\mathrm{ns}}{dt}=J^\mathrm{ns}_\mathrm{Q},
    \label{Eq-RelaxSpinNuclear}
\end{equation}
where $C_\mathrm{ns}$ is the heat capacity of nuclear spins at temperature $T_\mathrm{ns}$, and $J^\mathrm{ns}_\mathrm{Q}$ is the heat flow between nuclear spins and electrons. 
In the high-temperature approximation essentially valid throughout this experiment ($\mu B/k_\mathrm{B}T_\mathrm{ns}\ll1$), $C_\mathrm{ns}$ and $J^\mathrm{ns}_\mathrm{Q}$ reduce to simple expressions (see Appendix~C for full expressions, and Fig.\ \ref{fig-HighTempLim} for a test of the high-temperature approximation):
\begin{equation}
    C_\mathrm{ns}\simeq \gamma_\mathrm{ns} \Omega  B^2/T_\mathrm{ns}^2
   \label{Eq-CnslowBsT}
\end{equation}
and 
\begin{equation}
J^\mathrm{ns}_\mathrm{Q}(T,T_\mathrm{ns})\simeq\frac{C_\mathrm{ns} T_\mathrm{ns}}{K} (T-T_\mathrm{ns})\simeq\frac{\gamma_\mathrm{ns}\Omega B^2}{K T_\mathrm{ns}}(T-T_\mathrm{ns}),
\label{Eq-HeatFlowToSpins}
\end{equation}
with $\gamma_\mathrm{ns}$ the Curie constant of the nuclear spins. 
Electron heating dynamics remains determined by Eq.\ \eqref{Eq-ElectronDynamics}, but with the nuclear spin contribution $J^\mathrm{ns}_\mathrm{Q}$ now included in the total heat current emitted from the island electron bath:
\begin{equation}
J_\mathrm{Q}=J^\mathrm{el}_\mathrm{Q}(T,T_\mathrm{b})+J^\mathrm{ph}_\mathrm{Q}(T,T_\mathrm{b})+J^\mathrm{ns}_\mathrm{Q}(T,T_\mathrm{ns}).
\label{Eq-HeatBalance}
\end{equation}

In order to obtain $T$ and $T_\mathrm{ns}$, Eqs.~\eqref{Eq-ElectronDynamics} and \eqref{Eq-RelaxSpinNuclear} must be solved self-consistently.
Since the electron relaxation time is much shorter than the characteristic time for the nuclear spin equilibration ($K/T_\mathrm{ns}$), the electron heating dynamics can be solved to a good approximation via Eq.~\eqref{Eq-ElectronDynamics} by assuming a $T_\mathrm{ns}(t)$ that is fixed in time.
Moreover, at a 1\,s resolution, the experiment probes the electron temperature after the fast ($\lesssim10$\,ms) electron relaxation is achieved, which is given by solving the quasistationary relation $P_\mathrm{J}\simeq J_\mathrm{Q}(T,T_\mathrm{ns})$ (that does not involve the electronic heat capacity $C_\mathrm{e}$).
The resulting $T(T_\mathrm{ns})$ is then inserted into Eq.~\eqref{Eq-RelaxSpinNuclear} to obtain the slower evolution $T_\mathrm{ns}(t)$ and the associated $T(t)$.

For the thermal relaxation time $\tau$ toward $T_\mathrm{final}$, we find (Appendix D)
\begin{equation}
    \tau=\frac{C_\mathrm{ns}}{G_\mathrm{Q}^\mathrm{ns}}+\frac{C_\mathrm{ns}}{G_\mathrm{Q}^\mathrm{stat}},
    \label{Eq-TauGeneral}
\end{equation}
with the `stationary' heat conductance $G_\mathrm{Q}^\mathrm{stat}=G_\mathrm{Q}^\mathrm{el}+G_\mathrm{Q}^\mathrm{ph}$ including both the thermal conductances via electronic channels $G_\mathrm{Q}^\mathrm{el}=N_\mathrm{eff}\frac{\pi^2k_\mathrm{b}^2T}{3h}$ (with $N-1\leq N_\mathrm{eff}\leq N$ the effective number of electronic channels) and to phonons $G_\mathrm{Q}^\mathrm{ph}=\alpha \Sigma \Omega T^{\alpha-1}$ (with $4<\alpha<6$ a power law coefficient depending on disorder and $\Sigma$ the coupling strength), and with the thermal conductance $G_\mathrm{Q}^\mathrm{ns}$ between electrons and nuclear spins given for $\mu B/k_\mathrm{B}T\ll1$ by $G_\mathrm{Q}^\mathrm{ns}\simeq C_\mathrm{ns}T/K$. 

A quantitative comparison with the data therefore requires us to characterize $J^\mathrm{stat}_\mathrm{Q}=J^\mathrm{el}_\mathrm{Q}+J^\mathrm{ph}_\mathrm{Q}$.
This can be done separately from the nuclear spin contribution, by focusing on the stationary regime where $T=T_\mathrm{ns}$ and $J^\mathrm{ns}_\mathrm{Q}=0$.
In such a steady-state condition, $T$ is set by the stationary heat balance equation $P_\mathrm{J}=J_\mathrm{Q}^\mathrm{el}+J_\mathrm{Q}^\mathrm{ph}$.
The electronic heat flow contribution is given by $ J_\mathrm{Q}^\mathrm{el}=N_\mathrm{eff}\times J_\mathrm{K}$, where $N_\mathrm{eff}(T)\in[N-1,N]$ is an effective number of electronic heat channels. 
Because of heat Coulomb blockade taking place on mesoscopic islands, $N_\mathrm{eff}$ is reduced by at most one compared to the total number $N$ of opened electronic channels \cite{Slobodeniuk2013,Sivre2018} (see Appendix~E for a full analytical expression of $J_\mathrm{Q}^\mathrm{el}$).
The second, electron-phonon contribution reads $J^\mathrm{ph}_\mathrm{Q}=\Sigma \Omega (T^\alpha-T_\mathrm{b}^\alpha)$ in metals.
We find that typical values $\alpha=5.5$ and $\Sigma\approx2$\,nW.$\mu$m$^{-3}$.K$^{-5.5}$, together with a $T$-independent $N_\mathrm{eff}$, match the stationary heat flow data at measurement accuracy (see Fig.~\ref{fig-CharactJelJph} and Table~\ref{tab_FigJQstat} in Appendix~E).
In practice, the smallest island is always in the fully developed heat Coulomb blockade regime ($N_\mathrm{eff}\simeq N-1$), as predicted from its known charging energy, whereas heat Coulomb blockade can be suppressed for the larger island.

With the separately established expression for $J^\mathrm{stat}_\mathrm{Q}$ matching the stationary data, the only remaining parameters of the thermal model are the nuclear spin Curie constant $\gamma_\mathrm{ns}$ and Korringa coefficient $K$. 
The colored lines in Fig.~\ref{fig-TauVersusNandV} represent $\tau$ calculated from Eq.~\eqref{Eq-TauGeneral} for the same configurations as the symbols of corresponding colors.
Importantly, the data-theory comparison displayed in Fig.~\ref{fig-TauVersusNandV} involves only a single adjustable parameter, $K=0.27$\,s.K, whereas the Curie constant is here expressed as a function of the Korringa coefficient: As discussed in the next section, $\gamma_\mathrm{ns}/K=0.13$\,J.s$^{-1}$.T$^{-2}$.m$^{-3}$ is independently set from the relative amplitude of the slow temperature evolution.
Note that the prediction uncertainty resulting from those on $\gamma_\mathrm{ns}/K$, $T_\mathrm{b}$ and $G_\mathrm{Q}^\mathrm{stat}$ is displayed by the surrounding semitransparent areas (Appendix~B).
When the temperature amplitude of the slow stage is small, $\tau$ is more difficult to extract from the measurements (see e.g.\ Fig.~\ref{fig-Island} in Appendix~B for the most challenging case of the smallest island). If the standard error on $\tau$ is larger than symbol size, it is displayed as error bars.

Overall, we observe a quantitative match between a set of 54 $\tau$ data points spanning over a full order of magnitude, from 8\,s to 160\,s, and the prediction of single nuclear spin bath model involving $K$ as only adjustable parameter.
With a reasonable value of the Korringa coefficient ($K=0.27$\,s.K), predicted and extracted $\tau$ remain close while varying the initial and final temperatures, the number of channels, the magnetic field and the island's volume (see Appendix~H for supplemental data on a different sample).

\begin{figure}[htb]
\centering\includegraphics[width=1\columnwidth]{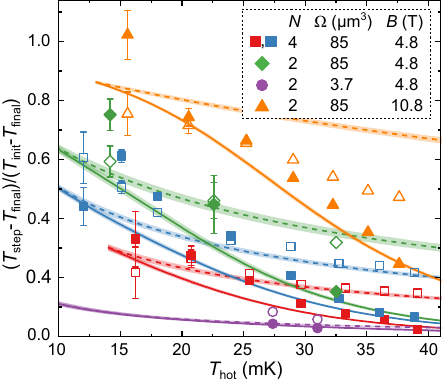}
\caption{
\footnotesize
Relative amplitude of the slow temperature evolution versus experimental parameters.
Symbols: data obtained from the same $T(t)$ measurements as $\tau$ in Fig.~\ref{fig-TauVersusNandV}, and plotted as a function of the high temperature $T_\mathrm{hot}$.
Note that the red data were measured at $T_\mathrm{b}=14$\,mK, the orange data at $T_\mathrm{b}=13$\,mK and those in blue, magenta and green at $T_\mathrm{b}=9$\,mK.
Full/open symbols correspond to a temperature step up/down ($T_\mathrm{hot}=T_\mathrm{final/init}$).
The statistical standard error is displayed as an error bar when larger than the symbol.
Lines: predictions of Eq.~\eqref{Eq-RelativeStepHeight} when heating (full lines, to be compared with full symbols) and cooling (dashed lines, to be compared with open symbols), for the same parameters as symbols of matching color.
The semitransparent broadening of predictions displays the standard uncertainty from the separately characterized $G_\mathrm{Q}^\mathrm{stat}$ and $T_\mathrm{b}$.
}
\label{fig-FullEvol}
\end{figure}

\section{RELATIVE AMPLITUDE OF FAST AND SLOW EVOLUTION}

Beyond the thermalization time, another characteristic of the data that can provide insight into the underlying mechanism is the relative temperature amplitude of the fast and slow thermal evolutions.
Figure~\ref{fig-FullEvol} shows extracted values of the relative height of the slow temperature evolution with respect to the total temperature change $|T_\mathrm{final}-T_\mathrm{init}|$. 
It is plotted as symbols for different configurations ($N$, $\Omega$, $T_\mathrm{init}$, $T_\mathrm{final}$, $B$) as a function of the hottest temperature $T_\mathrm{hot}$ ($T_\mathrm{hot}=T_\mathrm{final}$ when heating; $T_\mathrm{hot}=T_\mathrm{init}$ when cooling).
Note that the experimental uncertainty varies between data points. In particular, for smaller $|T_\mathrm{final}-T_\mathrm{init}|$, the experimental error on the displayed relative amplitude of the slow $T(t)$ evolution is generally larger. 
In Fig.~\ref{fig-FullEvol}, error bars show the statistical standard error when it is larger than the size of the symbol.
Remarkably, the relative amplitude of the slow temperature evolution is found to extend over the full range, depending on the implemented configuration.

Theoretically, we expect the relative amplitude of the two consecutive thermalization stages to follow:
\begin{equation}
    \frac{{T_\mathrm{step}}-{T_\mathrm{final}}}{T_\mathrm{init}-T_\mathrm{final}}\simeq\frac{G_\mathrm{Q}^\mathrm{ns}(T_\mathrm{init})}{
    G_\mathrm{Q}^\mathrm{stat}(T_\mathrm{final})}+G_\mathrm{Q}^\mathrm{ns}(T_\mathrm{init}),\label{Eq-RelativeStepHeight}
\end{equation}
within the approximations of small enough $1-T_\mathrm{step}/T_\mathrm{final}$ such that the stationary (electronic and electron-phonon) heat current $J_\mathrm{Q}^\mathrm{stat}$ is essentially linear in $T_\mathrm{step}-T_\mathrm{final}$
and of low enough $\mu B/k_\mathrm{B}T$ for Eq.~\eqref{Eq-HeatFlowToSpins} to apply (Appendix~D).
Here, $T_\mathrm{step}$ is the intermediate electron temperature shortly after the $P_\mathrm{J}$ step, and $T_\mathrm{init(final)}$ corresponds to the steady-state electron temperature before (after) the $P_\mathrm{J}$ step.
Note that $G_\mathrm{Q}^\mathrm{ns}$ is taken at the temperature $T_\mathrm{ns}=T_\mathrm{init}$, as the nuclear spin temperature is yet unchanged shortly after the fast $T$ step.
Importantly, the temperature amplitudes of the fast and slow thermal evolutions can be similar to each other in the mesoscopic regime.
What determines their relative importance is the thermal anchoring of the electrons to the nuclear spins compared to the other baths.
If the thermal conductance between electron and nuclear spin dominates (limit of large volume, high field, low temperature) then the fast temperature step becomes vanishingly small, since $T$ essentially sticks to the nuclear spin temperature.
This corresponds to the macroscopic regime of adiabatic demagnetization refrigerators.
In the opposite microscopic limit of a relatively small thermal conductance to the nuclear spins (low volume, low field, high temperature), the slow electronic temperature evolution becomes very small, since the nuclear spins can be ignored.
Whereas the smaller island is close to the microscopic limit, with a slow $T(t)$ evolution of relatively small amplitude, the larger island displays a slow evolution amplitude spanning most of the intermediate regime (see Fig.~\ref{fig-FullEvol}).

The predictions of Eq.~\eqref{Eq-RelativeStepHeight} for the relative amplitude of the slow temperature evolution are displayed as full and dashed lines in Fig.~\ref{fig-FullEvol} (for heating and cooling, respectively).
Importantly, the only relevant nuclear spin parameter is here the ratio $\gamma_\mathrm{ns}/K$ [from $G^\mathrm{ns}_\mathrm{Q}\simeq (\gamma_\mathrm{ns}/K) \Omega B^2/T_\mathrm{ns}$].
Using the separately characterized expressions of $G_\mathrm{Q}^\mathrm{stat}$, the only adjustable parameter of the thermal model for the data-theory comparison in Fig.~\ref{fig-FullEvol} is $\gamma_\mathrm{ns}/K=0.13$\ J.s$^{-1}$.T$^{-2}$.m$^{-3}$.
Comparing full (open) symbols with continuous (dashed) lines of matching color on the 54-point dataset, we find that the single-parameter model reproduces the observed relative amplitude evolution, from a few percent to nearly the full two-step temperature change $T_\mathrm{final}-T_\mathrm{init}$, upon spanning a variety of configurations.
The limited quantitative discrepancies, of up to about 0.1, are attributed to the simplification of the model, including the linearization of $J^\mathrm{el}_\mathrm{Q}$ and $J^\mathrm{ph}_\mathrm{Q}$ (see Appendix~F and Fig.~\ref{fig-SI-Tstep}), the low magnetic to thermal energy approximation (see Appendix~D) and the hypothesis of a single nuclear spin specie.
Importantly, the nuclear spin parameters $\gamma_\mathrm{ns}\simeq0.04$\,J.K.T$^{-2}$.m$^{-3}$ and $K\simeq0.3$\,s.K extracted from the data-theory comparison in Figs.\ref{fig-TauVersusNandV} and \ref{fig-FullEvol} are consistent with the tabulated values of the atoms composing the islands (see Table~I in Appendix~C).
Together, the quantitative data-theory comparisons in Figs.~\ref{fig-TauVersusNandV} and \ref{fig-FullEvol} (see also Fig.~\ref{fig-HighTempLim} for the full time evolution and Fig.~\ref{fig-SI-Tstep-intermediate} for additional data on a different device) attests to a thorough understanding of the observed two~-step heating dynamics of mesoscopic islands.

\section{CONCLUSION}

We have shown that mesoscopic metal islands at high magnetic fields and low temperatures display two-step heating and cooling dynamics reflecting a balanced competition between heat transfers with electron, phonon, and nuclear spin systems.
Remarkably, the effect of nuclear spins is not visible in the steady-state heat currents investigated in earlier work; they appear only in the heating dynamics due to their essentially local and exclusive coupling to electrons. 
The impact of nuclear spins will certainly play a role in mesoscopic thermodynamic investigations that have been proposed or are on the horizon, for example in measurements of the fluctuations of thermodynamic quantities, especially in the much-studied quantum Hall regime \cite{Ronetti_HOMheatnoise_2019,Battista_dJQ_2013,Dashti_dJQ_2018,SeifertThermoInference2019,Crepieux_dJQ_2021,Schiller_dJQ_2022,palmqvist2025combiningkineticthermodynamicuncertainty}.
The understanding developed in this work will immediately apply to the thermal engineering of nanodevices.
In particular, mesoscopic islands of adequate dimensions can exhibit a short thermal response time that allows for synchronous, high-precision measurements of central thermodynamic quantities in quantum Hall systems.
Those include thermoelectric properties, entropy or heat currents \cite{Chickering2010,Hartman2018}, which are considered particularly revealing observables of the exotic character of emergent states.
The observation and quantitative understanding of the two-step thermalization that is demonstrated here, therefore, establishes the nuclear spin bath as an essential element that must be considered for manipulating or exploiting heat in mesoscopic circuits.

\begin{acknowledgments} 
This work was supported by the European Research Council (ERC-2020-SyG-951451). 
This work was done within the C2N micro nanotechnologies platform and partly supported by the French RENATECH network and the general Council of Essonne.

We thank K.~Ensslin, T.~Ihn, Y.~Meir and E.~Sela for discussions, and H.~Bartolomei for his help on the measurement of the additional sample.

{\noindent\textbf{Author Contributions.}}
A.V., C.P.\ and F.Z.\ performed the experiment with input from A.Aa., A.An.\ and F.P.;
F.Z., A.V., C.P., A.An.\ and F.P.\ analyzed the data with input from J.F.;
C.P., A.V., Y.S.\ and A.Aa.\ fabricated the samples;
A.C.\ and U.G.\ grew the 2DEG;
Y.J.\ provided the HEMT for noise measurements;
F.Z., A.An.\ and F.P.\ wrote the paper with input from all authors;
A.An.\ and F.P.\ led the project.

{\noindent\textbf{Author Information.}}
The authors declare no competing financial interests.
Correspondence and requests for materials should be addressed to A.An.\ (anne.anthore@c2n.upsaclay.fr) and F.P.\ (frederic.pierre@cnrs.fr).

{\noindent\textbf{Data availability.}}
The data that support the findings of this article are openly available \cite{ZanichelliZenodo2026}.

\end{acknowledgments}

\section*{APPENDIX A: DEVICE, SETUP, TEMPERATURE}

\subsection*{1. Nanofabrication}
The sample discussed in the main text (see Appendix~H for the additional sample) is patterned by standard e-beam lithography on a GaAlAs heterostructure forming an electron gas 90\,nm below the surface, with a density of $2.6\times 10^{11}\,\mathrm{cm}^{-2}$ and a mobility of $0.5\times 10^{6}\,\mathrm{cm}^2V^{-1}\mathrm{s}^{-1}$. 
The 2DEG mesa is delimited by a wet etching approximately 100\,nm deep with a $\mathrm{H}_3\mathrm{PO}_4/\mathrm{H}_2\mathrm{O}_2/\mathrm{H}_2\mathrm{O}$ solution.
The ohmic contacts are realized with a metallic multilayer deposition of Ni(10\,nm)-Au(10\,nm)-Ge(90\,nm)-Ni(20\,nm)-Au(170\,nm)-Ni(40\,nm) followed by an annealing at 440°C where the metal penetrates the GaAlAs. 
The notched shape of the island in Fig.~\ref{fig-sample} increases the perimeter for an improved contact quality.
Note that the active outer quantum Hall edge channel is found to be perfectly connected to the small metallic island, at experimental accuracy (with a reflection probability below $0.3\%$).
This value is obtained from the ratio between the current reflected when the path to the island is barred and the current impinging on the same measurement electrode when the path to the island is open. See Methods in \cite{Iftikhar2015} for details on this self-calibrated procedure.
Note that the distance between the islands and the large electrodes is relatively important, of several hundred micrometers. In particular, the edge path from the smaller (bigger) metallic island to the large measurement electrode downstream is about 400\,$\mathrm{\mu}$m (550\,$\mathrm{\mu}$m) long.

\subsection*{2. Experimental setup}
The measurements are performed in a cryo-free dilution refrigerator with extensive measurement lines filtering and thermalization (see \cite{Iftikhar2016} for details).
Noise measurements for the thermometry described below are performed near 1\,MHz with homemade cryogenic amplifiers \cite{Jezouin2013b}.
The dc voltage source is realized by a dc current bias through a high resistance (100~M$\Omega$) driving a fixed sample resistance $h/\nu e^2$ to cold ground in parallel with a low temperature capacitance (100~nF), similarly to previous works investigating the stationary heat flow from small, heated-up islands in the quantum Hall regime. 
The corresponding $1/2\pi R C$ bandwidth of the injection lines used to dissipate Joule power in the mesoscopic islands is $\nu\times62$\,Hz.

\subsection*{3. Electronic temperature}
The electronic temperature $T$ is obtained from on-chip thermal noise measured on an ohmic contact.
For the $\nu=1$ data the conversion factor is calibrated from the linear slope of thermal noise vs temperature of the mixing chamber, at sufficiently high temperature where the difference between electron and mixing chamber temperatures is negligible.
In practice, the calibration is performed above 30\,mK where the high linearity of noise vs temperature attests to the good thermal anchoring of the electrons (here, differences between \textit{in-situ} electronic temperature $T$ and mixing chamber temperature develop essentially below 20\,mK). 
The corresponding data displayed in the main manuscript are obtained at $T_\mathrm{b}=13$\,mK.
For the $\nu=2$ data the conversion factor is calibrated from shot noise thermometry with a quantum point contact.
In both cases ($\nu=1$ and $\nu=2$), the standard fit error in the gain of the amplification chain is found to be below 1\%, and it is subsequently ignored. 
The corresponding data in the main manuscript are at $T_\mathrm{b}=9$ and 14\,mK. The uncertainty on the base temperature is estimated from the temperature drifts occurring during the measurements, which is of about $0.5$\,mK.

\section*{APPENDIX B: Fitting procedures}

\subsection*{1. Extraction of the thermalization time}

\begin{figure}[hb]
\centering\includegraphics[width=1.\columnwidth]{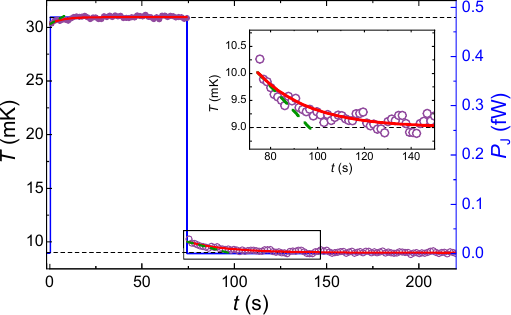}
\caption{
\footnotesize
Electronic temperature evolution of the small metallic island ($\Omega\simeq3.7\,\mu\mathrm{m}^3$, $N=2$ electronic channels, $T_\mathrm{b}=9$\,mK). 
Symbols are $T(t)$ measurements following a $P_\mathrm{J}$ step (see blue line and corresponding right axis), obtained by averaging $742$ individual traces. 
The size of the symbols matches the statistical standard error on the measurements.
Red lines are exponential fits with the characteristic time $\tau$ and the height of the slow temperature evolution as parameters.
The green dashed lines show the asymptotic slopes at short times.
Inset: zoom on the cooling dynamics.
}
\label{fig-Island}
\end{figure}

In this section, we detail the fitting procedure yielding the experimental slow thermalization time $\tau$ shown in Fig.~\ref{fig-TauVersusNandV} as symbols.
The fitted $T(t)$ data for a given configuration are obtained by averaging together multiple repetitions of $T(t)$ measurements, performing a first-order interpolation to account for small variations in the measurement times. 
For the larger island, we most often perform these $T(t)$ measurements on two nominally identical islands (showing identical behaviors) and average these data together.
For each sweep, corresponding to a step up or down in injected Joule power, the temperature difference $T-T_\mathrm{b}$ is obtained considering the excess noise with respect to the base temperature noise taken as the average noise of the last 25 points at $P_\mathrm{J}=0$.
Accordingly, the hot temperature $T_\mathrm{hot}$ is obtained from the mean noise of the last 25 points (before the Joule power is set back to zero to start a new sequence).
The response times $\tau$ are obtained by fitting the $T(t)$ data with the function $A\exp{-t/\tau}+T_\mathrm{final}$, where $T_\mathrm{final}$ (given by the average of the last 25 $T(t)$ data points before the next $P_\mathrm{J}$ step) corresponds to $T_\mathrm{b}$ or $T_\mathrm{hot}$.
As deviations to an exponential behavior develop away from $T_\mathrm{final}$, the fits (performed using the least square method) were restricted to data points at $t\geq t_\mathrm{min}$ with $t_\mathrm{min}$ the lowest time such that $|T(t_\mathrm{min})-T_\mathrm{final}|/T_\mathrm{final}<0.4$. 
In practice, this restriction essentially takes effect when the largest island is cooling to a base temperature of 9\,mK at $B=4.8$\,T or to 13\,mK at $B=10.8$\,T.
The error bars in Fig.~\ref{fig-TauVersusNandV} display the standard error of the fitting parameter $\tau$.
Illustrative fits are shown as red lines in Fig.~\ref{fig-ResponseTime} for the larger island where the relative amplitude of the slow evolution is larger, and in Fig.~\ref{fig-Island} for the small island where the relative amplitude can be correspondingly smaller. 
Note also that a sufficiently well-resolved slow stage amplitude is necessary to be able to extract $\tau$ from a $T(t)$ fit. 
In practice, we require that the amplitude of the slow thermalization stage is twice as big as the standard error on $T$.
This condition is always fulfilled for the data discussed in the main text; hence, the numbers of relative amplitude and $\tau$ data points in, respectively, Figs.~3 and 4 are identical. 
This condition was not always fulfilled for the $B=2.2$\,T and 4.5\,T configurations of the additional sample discussed in Appendix H. 
In particular, relative amplitude data points for heating up at $B=2.2$\,T in Fig.~\ref{fig-SI-Tstep-intermediate}(b) (full symbols) have no $\tau$ counterparts in Fig.~\ref{fig-SI-Tstep-intermediate}(a).

\subsection*{2. Extraction of the nuclear spin parameters}

\textit{a) Ratio $\gamma_\mathrm{ns}/K$.}
The ensemble of slow stage amplitude data points in Fig.~\ref{fig-FullEvol} is fitted using the least square method with individual weights associated with the standard error on the measured relative temperature step.
The fit function Eq.~\eqref{Eq-RelativeStepHeight} is written explicitly as a function of the relevant parameters in Eq.~\eqref{Eq-RelativeStepHeight-fitfn}.
The only adjustable parameter is here $\gamma_\mathrm{ns}/K$, the others being separately fixed.
We obtain $\gamma_\mathrm{ns}/K=0.13\pm0.01$\,J.s$^{-1}$.T$^{-2}$.m$^{-3}$, with the provided standard error including the propagated uncertainty on separately fixed parameters (see below).

\textit{b) Korringa coefficent $K$.}
The ensemble of $\tau$ data points in Fig.~\ref{fig-TauVersusNandV} is fitted using the least square method with individual weights associated with the standard error of the corresponding $\tau$.
The fit function obtained from Eq.~\eqref{Eq-TauGeneral} is written explicitly as a function of the relevant parameters in Eq.~\eqref{Eq-tau-fitfn}. 
The only adjustable parameter is here $K$.
Notably, the Curie constant coefficient $\gamma_\mathrm{ns}$ is set by the ratio $\gamma_\mathrm{ns}/K$ obtained from the separate fit of the slow stage amplitude discussed in \textit{a)}.
We obtain $K=0.27\pm0.02$\,s.K.

\textit{c) Standard errors.}
The provided standard error on the fitting parameters includes both the parameter's fit error and the contributions of the errors on all the separately fixed parameters, taking into account their covariance.
In practice, this is obtained by performing an ensemble of fits (10000) with the separately characterized parameters randomly distributed following Gaussian laws with the corresponding variance and covariances.
The standard uncertainty on the displayed predictions, shown as semitransparent areas around the theoretical lines, corresponds to the uncertainty from the separately fixed parameters.
In practice, the main source of error is found to be the uncertainty on $T_\mathrm{b}$, itself dominated by $\approx0.5$\,mK temperature drifts during the long measurements.

\textit{d) Alternative procedure.}
We compare the nuclear spin parameters extracted above in two steps with the result of a single global fit of all the data in Figs.~\ref{fig-TauVersusNandV} and \ref{fig-FullEvol}.
Fitting simultaneously the ensemble of $\tau$ and slow stage amplitude data using $K$ and $\gamma_\mathrm{ns}/K$ as two adjustable parameters, we obtain $\gamma_\mathrm{ns}/K=0.12\pm0.01$\,J.s$^{-1}$.T$^{-2}$.m$^{-3}$ and $K=0.29\pm0.02$\,s.K.
These values overlap within standard errors with those obtained using the main (two-step) fitting procedure detailed above.
Accordingly, the variance between best fit and data points is found identical, at the 1\% level, for both fitting strategies.
This confirms the robustness of our procedure.

\section*{APPENDIX C: Heat capacity and Korringa time for nuclear spins in ohmic contacts}

In this section, we first summarize the nuclear spin characteristics of the main constitutive elements of the islands. 
Then we recall the expressions of the heat capacity of nuclear spins and of the heat flow between electrons and nuclear spins beyond first order in $\mu B/k_\mathrm{B}T$.
In the third part, these expressions are used  to determine the influence of highest order terms in the experimental situations where $k_\mathrm{B}T\sim\mu B$.

\subsection*{1. Nuclear spin parameters}
In Table\,\ref{tab_ElementNScarac}, we provide important nuclear spin characteristics of the atoms deposited to realize the metallic islands and composing the semiconductor heterojunction. 
The displayed value of $|g_\mathrm{N}|\mu_\mathrm{N} B$ ($=\mu B/I$) at $B=10.8$\,T corresponds to the magnetic energy splitting between nuclear spin states, in mK, for each constitutive elements. Here $\mu_\mathrm{N}$ is the nuclear magneton, $g_\mathrm{N}$ the nuclear Landé factor and $I$ the nuclear spin. A comparison with the temperature allows one to straightforwardly evaluate the ratio of magnetic over thermal energy.
The Curie constant labeled $\gamma_\mathrm{ns}$ is given by Eq.~\eqref{Eq-CurieExpression}. 
At low enough magnetic to thermal energy, $\gamma_\mathrm{ns}$ is directly related to the heat capacity of the corresponding nuclear spin bath $C_\mathrm{ns}\simeq \gamma_\mathrm{ns}\Omega(B/T)^2$. 
Note that in the case of an alloy, the effective Curie constant per unit volume is the sum of the contributions from the constitutive materials weighted by the relative amount of this material with respect to the listed value for a pure material.

\noindent
\begin{table}[h!]
\begin{tabular}{|c|c|c|c|c|c|c|c|}
   \hline
    Element & Nuclear &$g_\mathrm{N}$ & 
     Atomic& {$|g_\mathrm{N}|\mu_\mathrm{N} B/k_\mathrm{B}$} & $\gamma_\mathrm{ns}$ \\
    &spin $I$&&density $\rho$&($B=10.8\,\mathrm{T}$)&\\
     & & &($10^{28}$m$^{-3}$)&(mK)&\(\left(\frac{\mathrm{J\,K}}{\mathrm{T}^{2}\,\mathrm{m}^{3}}\right)\) \\
     \hline
     $^{197}$Au & $3/2$ & $0.10$&  $5.9$&$0.39$&$0.0013$\\
     \hline
     $^{73}$Ge (8\%) & $9/2$& $-0.20$& $4.4$&$0.77$&$0.026$\\
     \hline
     $^{61}$Ni (1\%) & $3/2$&$-0.5$&$9.1$&$2.0$&$0.05$ \\ 
     \hline
     $^{69}$Ga (60\%) & $3/2$&$1.3$&$5.1$&$5.3$&$0.2$\\ 
     $^{71}$Ga (40\%)& $3/2$&$1.7$&$5.1$&$6.8$&$0.3$\\
     \hline
     $^{75}$As&$3/2$&$1.0$&$4.1$&$3.8$&$0.09$\\
     \hline
     $^{27}$Al&$5/2$&$1.5$&$6.0$&$5.8$&$0.7$\\
     \hline
\end{tabular}
\caption{Nuclear spin parameters of relevant elements. Displayed percentages indicate the natural fraction of the corresponding isotope (when different from 100\%). 
Isotopes with $I=0$ are not shown. 
The nuclear Landé factor is obtained from the value of $\mu=Ig_\mathrm{N}\mu_\mathrm{N}$ provided by WebElements.
The Curie constant $\gamma_\mathrm{ns}$ is calculated with Eq.~\eqref{Eq-CurieExpression}.} 
\label{tab_ElementNScarac}
\end{table}

The Korringa coefficient $K$ relates the nuclear spin thermal relaxation time in metals to the temperature through $\tau_\mathrm{ns}\simeq K/T_\mathrm{ns}$. 
Note that since in pure metals $K$ is inversely proportional to the square of the density of states at the Fermi energy \cite{Korringa1950}, a noticeable impact on $K$ is expected in alloys of modified density of states.
For the constitutive elements forming pure metallic phases Au, Al and Ni, typical values of $K$ observed at low temperatures are, respectively, 4.6\,s.K \cite{Narath1967}, 1.8\,s.K \cite{Anderson1959} and 0.2-2\,s.K \cite{Streever1973}.

\subsection*{2. Nuclear heat capacity and heat flow expressions at arbitrary $\mu B/k_\mathrm{B}T$}

The heat capacity $C_\mathrm{ns}$ of a nuclear spin bath results from the interaction between the nuclear magnetic moment $\mu$ of an atom of spin $I$ and the external magnetic field $B$. 
For a bath of nuclear spins $I$ at $T_\mathrm{ns}$, it reads \cite{Pobell2007}
\begin{equation}
\begin{split}
C_\mathrm{ns} =& \rho\Omega k_\mathrm{B} \left(\frac{g_\mathrm{N}\mu_\mathrm{N} B}{2k_\mathrm{B}T_\mathrm{ns}}\right)^2\left\{\sinh^{-2}{\left[\frac{g_\mathrm{N}\mu_\mathrm{N}B}{2k_\mathrm{B}T_\mathrm{ns}}\right]}\right.\\
&\left.-\left(2I+1\right)^2\sinh^{-2}{\left[\left(2I+1\right)\frac{g_\mathrm{N}\mu_\mathrm{N}B}{2k_\mathrm{B}T_\mathrm{ns}}\right]}\right\},
\label{Eq-CNSlowT}
\end{split}
\end{equation}
\noindent with $\rho$ the atomic density and $\Omega$ the volume.
At the smallest order in $\mu B/k_\mathrm{B} T_\mathrm{ns} \ll 1$, one recovers Eq.~\eqref{Eq-CnslowBsT} with the Curie constant given by 
\begin{equation}
    \gamma_\mathrm{ns}=\frac{\rho  I\left(I+1\right)g_\mathrm{N}^2\mu_\mathrm{N}^2}{3 k_\mathrm{B}}.
    \label{Eq-CurieExpression}
\end{equation}

The heat current $J^\mathrm{ns}_\mathrm{Q}$ results from the hyperfine interaction between nuclear spins and electron spins.
Considering only the so-called contact interaction and using the Fermi golden rule in the simpler case of a nuclear spin $I=\frac{1}{2}$, one gets the standard expression:
\begin{equation}
   J^\mathrm{ns}_\mathrm{Q}
  = -\rho\Omega M g_\mathrm{N}^2\mu_\mathrm{N}^2B^2
   \frac{\sinh{\left[\frac{g_\mathrm{N}\mu_\mathrm{N}B}{2k_\mathrm{B}}\left(\frac{1}{T} - \frac{1}{T_\mathrm{ns}}\right)\right]}}{2 \cosh{\left[\frac{g_\mathrm{N}\mu_\mathrm{N}B}{2 k_\mathrm{B}T_\mathrm{ns}}\right]} \sinh{\left[\frac{g_\mathrm{N}\mu_\mathrm{N}B}{2 k_\mathrm{B}T}\right]}},
  \label{Eq-JNSlowT}
\end{equation}
\noindent with $M$ a coupling constant accounting for the interaction amplitude between electrons and nuclear spins. For a generalization to arbitrary $I$, see \cite{Pobell2007,Bacon1972}.
At smallest order in $\mu B/k_\mathrm{B} T_\mathrm{ns} \ll 1$, this expression reduces to Eq.~\eqref{Eq-HeatFlowToSpins} with the Korringa coefficient given at $I=1/2$ by $K=(2k_\mathrm{B}M)^{-1}$.

\subsection*{3. Test of high-temperature approximation}

In this section, we test whether, and at what accuracy, the use of predictions derived within the approximation $\mu B/k_\mathrm{B}T_\mathrm{ns}\ll1$ remains justified even when $\mu B\sim k_\mathrm{B}T_\mathrm{ns}$. 
As seen from the magnetic energy given in Table~\ref{tab_ElementNScarac} for the highest B=10.8\,T, the possible constitutive elements Ga, As and Al could approach this regime.

\begin{figure}[htb]
\centering\includegraphics[width=\columnwidth]{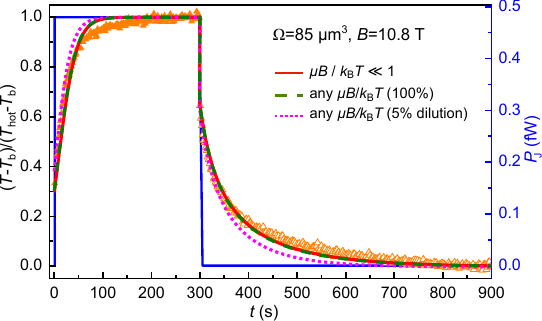}
\caption{
\footnotesize
Test of the high-temperature approximation on the full temperature evolution. 
Symbols represent the electronic temperature normalized by the full temperature increase when heating the large island from $T_\mathrm{b}=13$\,mK to $T_\mathrm{hot}=23.5$\,mK at $N=2$, $B=10.8$\,T due to  a change of injected Joule power (blue line, right axis).
It is compared to the predicted evolution using for $J_\mathrm{Q}^\mathrm{ns}$ and $C_\mathrm{ns}$ the equations \eqref{Eq-CNSlowT} and \eqref{Eq-JNSlowT} valid at arbitrary $\mu B/k_\mathrm{B}T_\mathrm{ns}$ for $I=1/2$ (green dashed line for a pure material, magenta dotted line for single active specie at a 5\% dilution), or Eqs.~\eqref{Eq-CnslowBsT} and \eqref{Eq-HeatFlowToSpins} derived assuming $\mu B/k_\mathrm{B}T_\mathrm{ns}\ll1$ (red full lines). 
\normalsize
}
\label{fig-HighTempLim}
\end{figure}
\normalsize

In Fig.~\ref{fig-HighTempLim}, we compare the calculation of the electron temperature evolution solving concomitantly Eqs.~\eqref{Eq-ElectronDynamics} and \eqref{Eq-RelaxSpinNuclear} using for $J_\mathrm{Q}^\mathrm{ns}$ and $C_\mathrm{ns}$ either the expressions of Eqs.~\eqref{Eq-CNSlowT} and \eqref{Eq-JNSlowT} valid at arbitrary $\mu B/k_\mathrm{B}T_\mathrm{ns}$ and using $I=1/2$ (dashed green lines) or Eqs.\eqref{Eq-CnslowBsT} and \eqref{Eq-HeatFlowToSpins} derived assuming $\mu B/k_\mathrm{B}T_\mathrm{ns}\ll1$ (red full lines).
The comparison is performed at $T_\mathrm{b}=13$\,mK and $B=10.8$\,T where $\mu_\mathrm{N}B/k_\mathrm{B}T_\mathrm{b}\simeq0.3$ is highest.
This comparison is displayed together with the corresponding data for the largest island.
The nuclear Landé factor is required for the full $I=1/2$ solution. 
We use $g_\mathrm{N}\simeq1.2$ estimated from $g_\mathrm{N}= \sqrt{4 k_\mathrm{B}\gamma_\mathrm{ns}/\rho}/\mu_\mathrm{N}$ with $\rho=6\times10^{28}$\,m$^{-3}$ the average density of atoms.
Despite a value of $g_\mathrm{N}$ close to the highest among constitute elements ($g_\mathrm{N}\simeq1.5$ for Al), we find that the experimental two-step behavior is equally well reproduced by both predictions, and that the difference between the high-temperature approximation and the full $I=1/2$ prediction remains invisible, well below the linewidth.

As a strongest possible test, we now assume that the dominant nuclear-spin material is a fraction of the island's constituent.
Among the possibly relevant atoms recapitulated in Table~\ref{tab_ElementNScarac}, the largest Curie constant (of Al) corresponds, when diluted at 5\%, to the fitted value $\gamma_\mathrm{ns}\simeq0.035=5\%\times0.7$\,J.K.T$^{-2}$.m$^{-3}$. 
In the $I=1/2$ model used for this test, the associated (highest possible) nuclear Landé factor becomes $g_\mathrm{N}\simeq5$, resulting in a relatively high magnetic energy of $k_\mathrm{B}\times20$\,mK at B=10.8\,T.
The corresponding full $T(t)$ predictions (magenta dotted line) is confronted in  Fig.~\ref{fig-HighTempLim} with the small magnetic energy approximation, at the highest magnetic to thermal energy ratio investigated experimentally.
In contrast with the nondilute case considered above, although the small and arbitrary magnetic energy predictions remain close, they are here noticeably different. 
The specifically investigated relative amplitude and characteristic time of the slow evolution are found to be reduced by at most 30\% for $\tau$ (for cooling, 20\% for heating) and 11\% for the amplitude (for heating, 3\% for cooling), at the highest magnetic to thermal energy ratio considered in Fig.~\ref{fig-HighTempLim}.

For a more complete assessment of the low magnetic energy approximation, we determine the change in the best fit values of the nuclear spin parameters within the most stringent case of the smallest possible fraction of active nuclear spins.
Performing a fully nonlinear global fit of all the data displayed in Figs.~\ref{fig-TauVersusNandV} and \ref{fig-FullEvol} (the sequential approach of two single-parameter fits used for low magnetic energies does not apply here), we observe a similar data-prediction agreement as with the nonlinear fit within the low magnetic energy approximation  (slightly improved, with a 20\% reduction of the variance).
The best fit nuclear spin parameters are found to be $\gamma_\mathrm{ns}/K\simeq=0.17\pm0.01$\,J.s$^{-1}$.T$^{-2}$.m$^{-3}$ and $K\simeq0.30\pm0.02$\,K.s, relatively similar to the values obtained within the low magnetic energy approximation. 
Note that the (smallest possible) dilution, associated with $\gamma_\mathrm{ns}\simeq0.05\simeq7\%\times0.7$\,J.K.T$^{-2}$.m$^{-3}$ corresponding to the best global fit parameters, is here of 7\%.

In summary, even for the most stringent dilute case and in the explored configuration of highest $\mu_\mathrm{N}B/k_\mathrm{B}T_\mathrm{b}$, the discrepancy with the high-temperature approximation remains moderate.
Furthermore, the best fit agreement between full data set and predictions is preserved. 
This robustness legitimates the use of the high-temperature approximation.

\section*{APPENDIX D: THERMAL DYNAMICS PREDICTIONS}

In this section, we first derive the intermediate electronic temperature $T_\mathrm{step}$, shortly after the application of an abrupt change in injected power.
This is done within the high-temperature approximation for nuclear spins and assuming that $T_\mathrm{step}-T_\mathrm{final}$ is small enough to allow for a linearization of the electronic and electron-phonon heat flows.
We then obtain the predicted characteristic thermalization time $\tau$ for the slow evolution stage, within the high-temperature approximation. 
The lumped-element thermal model is displayed in the inset in Fig.~\ref{fig-sample}.

\textit{1) Relative amplitude.}
The intermediate temperature $T_\mathrm{step}$ is obtained by solving
\begin{equation}
P_\mathrm{J}=J_\mathrm{Q}^\mathrm{stat}(T_\mathrm{step},T_\mathrm{b})+J_\mathrm{Q}^\mathrm{ns}(T_\mathrm{step},T_\mathrm{init}).
\label{Eq-intermediate-T-solution}
\end{equation}
The quasistationary contribution corresponds in our device to the sum of the electronic and electron-phonon heat currents: \begin{equation}J_\mathrm{Q}^\mathrm{stat}(T,T_\mathrm{b})=N_\mathrm{eff}\frac{\pi^2 k_\mathrm{B}^2}{6h}(T^2-T_\mathrm{b}^2)+\Sigma \Omega (T^\alpha-T_\mathrm{b}^\alpha)\label{eq:effective-stat-heat-flow}\end{equation}
The electron to nuclear spin heat current reads, for small enough magnetic energy $$J_\mathrm{Q}^\mathrm{ns}(T,T_\mathrm{ns})\simeq\frac{\gamma_\mathrm{ns}\Omega B^2}{K T_\mathrm{ns}}(T-T_\mathrm{ns})=G_\mathrm{Q}^\mathrm{ns}(T_\mathrm{ns})\times(T-T_\mathrm{ns}).$$
The relative amplitude of the slow evolution stage also requires the knowledge of $T_\mathrm{final}$, which is given by
\begin{equation}
   P_\mathrm{J}=J_\mathrm{Q}^\mathrm{stat}(T_\mathrm{final},T_\mathrm{b}).
   \label{Eq-final-T-solution}
\end{equation}
From the identical $P_\mathrm{J}$ in Eqs.\,\eqref{Eq-intermediate-T-solution} and \eqref{Eq-final-T-solution}, we have $$J_\mathrm{Q}^\mathrm{stat}(T_\mathrm{step},T_\mathrm{b})+G_\mathrm{Q}^\mathrm{ns}(T_\mathrm{ns})\times(T-T_\mathrm{ns})=J_\mathrm{Q}^\mathrm{stat}(T_\mathrm{final},T_\mathrm{b}).$$
For small enough $T_\mathrm{step}-T_\mathrm{final}$, we can make the linear approximation 
$J_\mathrm{Q}^\mathrm{stat}(T_\mathrm{final},T_\mathrm{b})-J_\mathrm{Q}^\mathrm{stat}(T_\mathrm{step},T_\mathrm{b})\simeq G_\mathrm{Q}^\mathrm{stat}(T_\mathrm{final})\times(T_\mathrm{final}-T_\mathrm{step})$
in the above equality, which allows us to recover Eq.~\eqref{Eq-RelativeStepHeight}.
Written explicitly in terms of the characterization parameters,  Eq.~\eqref{Eq-RelativeStepHeight} becomes
\begin{equation}
\begin{split}
     &\frac{{T_\mathrm{step}}-{T_\mathrm{final}}}{T_\mathrm{init}-T_\mathrm{final}}\simeq\\
     &\frac{\gamma_\mathrm{ns} \Omega B^2/K T_\mathrm{init}}{N_\mathrm{eff}\pi^2k_\mathrm{b}^2T_\mathrm{final}/3h+\alpha \Sigma \Omega T_\mathrm{final}^{\alpha-1}+\gamma_\mathrm{ns} \Omega B^2/K T_\mathrm{init}}.\label{Eq-RelativeStepHeight-fitfn}
\end{split}
\end{equation}

\textit{2) Thermalization time.}
We now turn to the time evolution during the second slow change.
In this stage, the electron temperature $T$ evolves from $T_\mathrm{step}$ to $T_\mathrm{final}$ as a function of $T_\mathrm{ns}$ according to the quasistationary heat balance: 
\begin{equation}
    P_\mathrm{J}=J_\mathrm{Q}^\mathrm{stat}(T,T_\mathrm{b})+G_\mathrm{Q}^\mathrm{ns}(T_\mathrm{ns})\times(T-T_\mathrm{ns}).
    \label{Eq-Thermalization_Appendix_intermediate1}
\end{equation}
Equations~\eqref{Eq-Thermalization_Appendix_intermediate1} and \eqref{Eq-final-T-solution} involve both the same $P_\mathrm{J}$, which gives
$$T-T_\mathrm{ns}=\frac{J_\mathrm{Q}^\mathrm{stat}(T_\mathrm{final},T_\mathrm{b})-J_\mathrm{Q}^\mathrm{stat}(T,T_\mathrm{b})}{G_\mathrm{Q}^\mathrm{ns}(T_\mathrm{ns})}.$$
Using this expression of $T-T_\mathrm{ns}$ in the temperature evolution of $T_\mathrm{ns}$ described by $C_\mathrm{ns}\, dT_\mathrm{ns}/dt=G_\mathrm{Q}^\mathrm{ns}\left(T-T_\mathrm{ns}\right)$, we obtain at first order in $T-T_\mathrm{final}$,
$$\frac{dT}{dt}+\frac{T-T_\mathrm{final}}{\tau}=0,$$
with the characteristic time $\tau$ given Eq.~\eqref{Eq-TauGeneral}. 
Explicitly in terms of the characterization parameters, $\tau$ reads
\begin{equation}
    \tau=\frac{K}{T_\mathrm{final}}+\frac{\gamma_\mathrm{ns} \Omega B^2}{N_\mathrm{eff} \pi^2 k_\mathrm{B}^2/(3h)\, T_\mathrm{final}^3+\alpha \Sigma \Omega\, T_\mathrm{final}^{\alpha+1}}.
    \label{Eq-tau-fitfn}
\end{equation}\\

\section*{APPENDIX E: Characterization of the stationary heat flow}

We describe here the procedure to characterize the stationary heat current in the steady-state regime, which includes the heat currents from the metallic island toward the phonon bath and through the electronic channels.
Note that the precise agreement demonstrated below (and previously observed in similar systems) between measured stationary heat flows and quantitative theoretical predictions attests that possible complications such as the cooling of electrons over long distances along the edge have a negligible impact on the heat balance of the island (as expected from the protection provided by quantum Hall chirality).

The characterization measurements are obtained by waiting enough time after changing the injected Joule power to reach the stationary regime (typically about 2\,min), such that nuclear spins and electrons are at the same temperature ($T\simeq T_\mathrm{ns}\simeq T_\mathrm{final}$, $J_\mathrm{Q}^\mathrm{ns}\simeq0$).
The steady-state heat balance then reads
\begin{equation}
P_\mathrm{J}=J^\mathrm{el}_\mathrm{Q}(T,T_\mathrm{b})+J^\mathrm{ph}_\mathrm{Q}(T,T_\mathrm{b}),
\label{Eq-HeatBalance2}
\end{equation}
where $J^\mathrm{ph}_\mathrm{Q}=\Sigma \Omega (T^\alpha-T_\mathrm{b}^\alpha)$ with $\Sigma$  the coupling strength, $\Omega$ the volume of the island and $4<\alpha<6$ a power law coefficient that depends on disorder.
From \cite{Sivre2018,Slobodeniuk2013}, the analytical expression $J^\mathrm{el}_\mathrm{Q}$ for the electronic heat flow through $N$ ballistic electronic channels connecting a metallic island of charging energy $E_\mathrm{C}$ and electronic temperature $T$, to large electrodes at temperature $T_\mathrm{b}$ is
\begin{equation}
\begin{split}
J_\mathrm{thy}^\mathrm{el}(N,T,&T_\mathrm{b},E_\mathrm{C})  = N\frac{\pi^2k_\mathrm{B}^2}{6h}(T^2-T_\mathrm{b}^2)\\
&+\frac{N^2E_\mathrm{C}^2}{\pi^2h}\left[I\left( \frac{NE_\mathrm{C}}{\pi k_\mathrm{B}T_\mathrm{b}}  \right) -I\left( \frac{NE_\mathrm{C}}{\pi k_\mathrm{B}T} \right) \right], \label{eqSLS}
\end{split}
\end{equation}
with the function $I$ given by
\begin{equation}
I(x)=\frac{1}{2}\left[ \ln\left(\frac{x}{2\pi}\right) - \frac{\pi}{x} -\psi\left( \frac{x}{2\pi} \right) \right],\label{eqI}
\end{equation}
where $\psi(z)$ is the digamma function.

\begin{figure*}[htb]
\centering\includegraphics[width=0.8\textwidth]{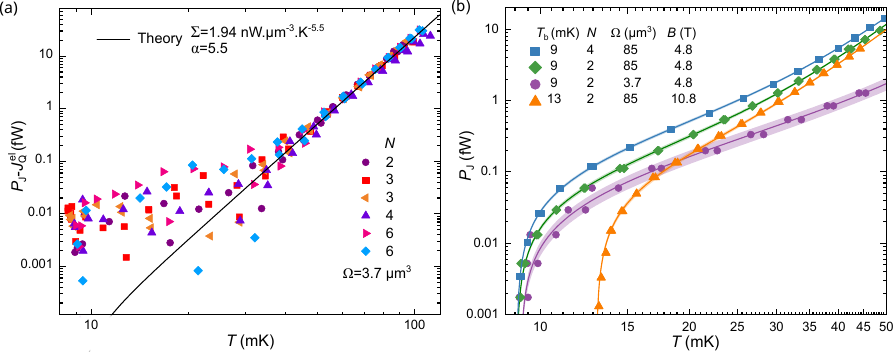}
\caption{
\footnotesize
Stationary heat current characterization.
(a) The electron-phonon heat flow is revealed by subtracting the predicted heat Coulomb blockade electronic contribution for the small island of known $E_\mathrm{C}$. 
The remaining $P_\mathrm{J}-J_\mathrm{Q}^\mathrm{el}$ (symbols) collapse onto each other independently of $N\in\{2,3,4,6\}$, as expected for $J_\mathrm{Q}^\mathrm{ph}$.
The line displays the electron-phonon predictions adjusted to match the $J_\mathrm{Q}^\mathrm{stat}$ data for $N=2$ at $T<50$\,mK ($\alpha=5.5$, $\Sigma\simeq1.9$\,nW.$\mu$m$^{-3}$.K$^{-5.5}$). 
(b) Best fits of the measured stationary heat current $J_\mathrm{Q}^\mathrm{stat}=P_\mathrm{J}$ performed in the configurations and temperature ranges where the thermal dynamics is investigated.
Semitransparent areas represent the standard error on the fitting parameters (see Table~\ref{tab_FigJQstat}), including covariance between $N_\mathrm{eff}$ and $\Sigma$.
\normalsize}
\label{fig-CharactJelJph}
\end{figure*}

For the small metallic island, the charging energy $E_\mathrm{C}=34.0\pm0.2\,\mu e$V is obtained from Coulomb diamonds measured in a single-electron transistor configuration.
The temperature $T$ of the island versus the injected Joule power $P_\mathrm{J}$ is measured for $2\leq N\leq 6$ connected ballistic electronic channels.
In Fig.~\ref{fig-CharactJelJph}(a), we display the total stationary heat flow from which we subtracted the predicted quantitative prediction for the electronic heat flow.
In the absence of any free parameters, since $E_\mathrm{C}$ is known for the small island, we find that all the data points for different $N$ and $T$ collapse on each other. 
This attests that the $N$-dependent contribution to the measured $J_\mathrm{Q}^\mathrm{stat}$ closely matches the predicted $J_\mathrm{Q}^\mathrm{el}$, without any discernible complications. 
The continuous line displays the $N$-independent prediction for electron-phonon using the typical parameters $\alpha=5.5$ and $\Sigma=1.94$\,nW.$\mu$m$^{-3}$.K$^{-5.5}$ with $\Omega=3.74\,\mathrm{\mu m}^3$ the geometric volume of the deposited material composing the island.
Note that the discrepancies developing at low temperatures are made more visible in log-log scale, but are not significant because they correspond to a small fraction of the subtracted electronic heat flow.

We then turn to individual $J_\mathrm{Q}^\mathrm{stat}$ calibration. 
Figure~\ref{fig-CharactJelJph}(b) shows, as symbols of shape and color identical to those in the main text, the measured $J_\mathrm{Q}^\mathrm{stat}$ in every configuration ($N$, $B$, $\Omega$) used to investigate the heating dynamics, and over the same range of temperature ($T<50$\,mK).
Lines of matching colors represent the fits of these data with Eq.\eqref{eq:effective-stat-heat-flow}, using $N_\mathrm{eff}$ and $\Sigma$ as adjustable parameters and $\alpha=5.5$ (see Table~\ref{tab_FigJQstat}).
The precise match between the measured and fitted $J_\mathrm{Q}^\mathrm{stat}$ ascertains that using the corresponding values of $\alpha$, $\Sigma$ and $N_\mathrm{eff}$ in Eqs.~\eqref{Eq-tau-fitfn} and \eqref{Eq-RelativeStepHeight-fitfn} allows for precise heat dynamics predictions.

\noindent
\begin{table}[h!]
\begin{tabular}{|c|c|c|c|c|c|c|c|}
   \hline
    Configuration & $\alpha$ & $\Sigma$ & $N_\mathrm{eff}$\\
    $N$/$B$(T)/$\Omega$($\mu$m$^3$) & (fixed) & (nW.$\mu$m$^{-3}$.K$^{-5.5}$)&\\
     \hline
     4/4.8/85 & 5.5 & 1.93$\pm0.03$ & 2.90$\pm0.03$\\
     \hline
     2/4.8/85 & 5.5 & 1.65$\pm0.02$& 1.67$\pm0.02$\\
     \hline
     2/4.8/3.7 & 5.5 & 1.94$\pm0.95$ & 1.07$\pm0.05$\\
     \hline
     2/10.8/85 & 5.5 & 1.45$\pm0.02$ & 1.13$\pm0.01$\\
     \hline
\end{tabular}
\caption{Electronic and electron-phonon heat flow fit parameters used to match the measured $J_\mathrm{Q}^\mathrm{stat}$ at $T<50$\,mK. The large fit uncertainty on $\Sigma$ for the small island reflects the weaker influence of electron-phonon interactions in smaller volumes.} 
\label{tab_FigJQstat}
\end{table}

\section*{APPENDIX F: Slow evolution amplitude beyond small temperature differences}
In this section, we focus on the relative amplitude of the slow thermal relaxation, generalizing the theoretical treatment beyond the linear approximation ($1-T_\mathrm{step}/T_\mathrm{final}\ll1$) in Appendix~D.

\begin{figure}[htb]
\centering\includegraphics[width=1\columnwidth]{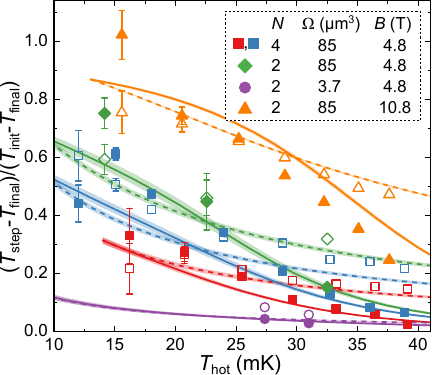}
\caption{
\footnotesize
Relative amplitude of the slow temperature evolution, beyond the small temperature difference approximation.
Symbols: the same data as in Fig.~\ref{fig-FullEvol}.
Lines: fit with the numerical solution of Eq.~\eqref{Eq-intermediate-T-solution} (continuous for heating to $T_\mathrm{hot}$ and dashed for cooling from $T_\mathrm{hot}$).
}
\label{fig-SI-Tstep}
\end{figure}

To extend the result of Eq.~\eqref{Eq-RelativeStepHeight}, we calculate the intermediate temperature $T_\mathrm{step}$ by solving numerically the electronic heat balance in Eq.~\eqref{Eq-intermediate-T-solution}.
The result for the relative height of the slow evolution is shown in Fig.~\ref{fig-SI-Tstep}, with continuous (dashed) lines corresponding to heating (cooling). 
The contribution of higher-order terms in temperature differences for the electronic, electron-phonon and electron-nuclear spin heat currents results in changes in the nonlinear prediction with respect to the linearized one that generally remains small, although it can reach about 30\% in the worst situations.
The overall agreement of the data with the nonlinear predictions is similar (noticeably better) than with the linear ones (the data-prediction variance is reduced by a factor of 2). 
The nonlinear predictions displayed in Fig.~\ref{fig-SI-Tstep} are obtained using the best nuclear spin parameter $\gamma_\mathrm{ns}/K=0.14\pm0.01$\,J.s$^{-1}$.T$^{-2}$.m$^{-3}$ specifically adjusted for these predictions.
Note that the difference with the best fit value obtained using linear predictions is within the standard uncertainties. 
Accordingly, this small change in $\gamma_\mathrm{ns}/K$ does not significantly impact the data-prediction comparison on the characteristic thermalization time $\tau$: We extract a best fit value $K=0.26\pm0.02$\,s.K, within standard error of the Korringa coefficient obtained using linear predictions.

\section*{APPENDIX G: Experimental tests}
The unanticipated character of the presently observed two-step thermalization process calls for a discussion of potential alternative explanations, such as a slow response of some other part of the system including the measurement setup or along the quantum Hall edges. 
We review here some of the strong constraints imposed by the body of observations when combined with further details on the measurement process and with additional tests.

\begin{figure*}[htb]
\centering\includegraphics[width=0.8\textwidth]{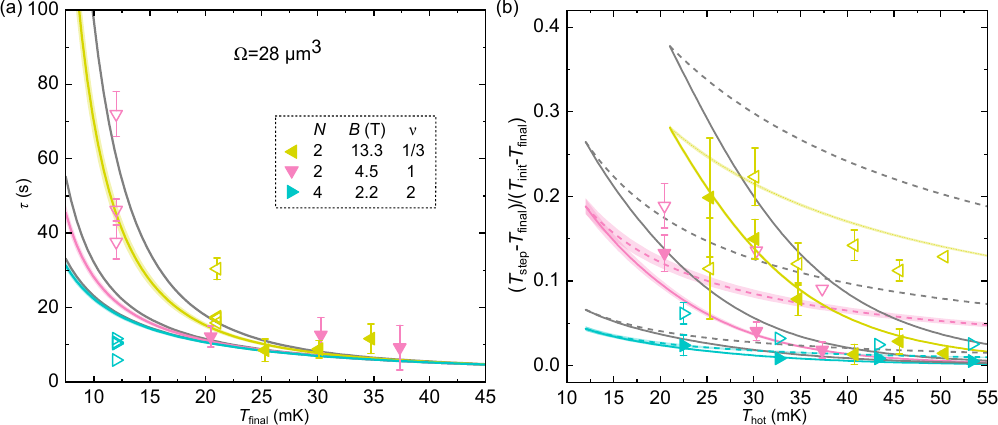}
\caption{
\footnotesize
Slow temperature evolution of additional device.
Data points while heating/cooling are shown as full/open symbols.
Colored lines are fits using $\gamma_\mathrm{ns}/K\simeq0.09$\,J.s$^{-1}$.T$^{-2}$.m$^{-3}$ and $K\simeq0.2$\,K.s in Eqs.~\eqref{Eq-RelativeStepHeight} and \eqref{Eq-TauGeneral}.
Gray lines are predictions using the nuclear spin parameters used for the sample discussed in the main text ($\gamma_\mathrm{ns}/K\simeq0.13$\,J.s$^{-1}$.T$^{-2}$.m$^{-3}$, $K\simeq0.3$\,K.s).
(a) Characteristic time $\tau$.
(b) Relative amplitude.
Full (dashed) lines correspond to heating (cooling) and should be compared to full (open) symbols of matching color. 
}
\label{fig-SI-Tstep-intermediate}
\end{figure*}

\textit{1) Ruling out instrumental bandwidth and filtering artifacts.}
The characteristic time and relative amplitude of the slow thermalization stage depend strongly on the island hot temperature (Figs.~\ref{fig-TauVersusNandV} and \ref{fig-FullEvol}).
Most of the data points displayed with identical symbols are obtained sequentially, for the same base temperature, with the same measurement protocol, and using the same setup.
The relatively small amplitude and short times at the highest island temperatures hence provide, very straightforwardly, maximum bounds for such measurement artifacts.
This corroborates the known instrumental and filtering bandwidth, several orders of magnitude larger than required for the temporal resolution of $\sim1$\,s (which is set by the noise integration time chosen to achieve the targeted temperature resolution).

\textit{2) Ruling out thermal transients along measurement paths} (including measurement lines and quantum Hall edge paths).
Heating up/cooling down an island between the same base temperature and similar $T_\mathrm{hot}$, we find a strong reduction of the amplitude and characteristic time of the slow response for smaller islands (see violet circles vs green lozenges in Figs.~\ref{fig-TauVersusNandV} and \ref{fig-FullEvol}).
These data were measured using the same electrical lines and instruments (with gates controlling whether the small island is connected), which allows us to rule these out as a possible explanation of the observed changes.
Furthermore, the edge path between small island and measurement pad is shared with the edge path originating from the largest island, corresponding to 75\% of the latter (400\,$\mu$m shared on a total of 530\,$\mu$m). 
This straightforwardly implies that any complications (unexpected and not detected in the stationary regime) distributed evenly along these paths should have a relative impact below 25\% on the slow thermalization stage when comparing these two configurations. 
This is in contrast with the relatively large changes observed, allowing us to rule out such a possibility.

\textit{3) Ruling out (transient and stationary) global chip heating.}
Two nominally identical implementation of the largest islands located 200\,$\mu$m away on the same chip (one shown in Fig.~\ref{fig-sample}) were simultaneously measured through different noise measurement lines, and found to show identical thermal behaviors both in the stationary regime and on the dynamical response on present interest. 
Applying the dissipated Joule power step on only one of these two islands, we could ascertain the absence of global chip heating (either transient or stationary) from the absence of any discernible change in the time-resolved noise signal originating from the other, unheated island (tests performed at $\nu=1$ and $\nu=2$; data not shown).

\textit{4) Ascertaining the island origin of transient noise.}
We checked that the applied voltage step normally used to dissipate power into the island does not induce any change in the time-resolved noise if the edge current is reflected just before reaching the island (using the gate controlling $N$; data not shown). This ascertains that the observed noise signal is a consequence of heating up specifically the mesoscopic islands.

\textit{5) Robustness vs sample specifics.}
We present in Appendix~H a set of complementary data that confirms our observations on a different sample based on a 2DEG of twice smaller density, with an island of intermediate size and over an extended range of magnetic fields up to the fractional filling factor $\nu=1/3$.

\section*{APPENDIX H: Additional device}
The present observation of a two-step thermalization is ascertained using a second sample, by measuring over an extended range of magnetic field a mesoscopic island of intermediate size, connected to a 2DEG of lower density.

\textit{1) Sample details.} 
The GaAlAs 2DEG material of the additional sample is deeper (140\,nm below the surface), of lower electronic density ($1.2\times10^{11}$\,cm$^{-2}$), and higher mobility ($1.8\times10^6$\,cm$^2$V$^{-1}$s$^{-1}$). 
Two nominally identical metallic island were fabricated on the same chip using identical fractions of Au, Ge and Ni (similarly annealed at 440°C) as for the islands studied in the main text but now deposited following a different stack Ni(10\,nm)-Ge(90\,nm)-Au(180\,nm)-Ni(60\,nm).
The nominal island volume of $\Omega\simeq28\,\mu\mathrm{m}^3$ is intermediate between the smallest and largest islands discussed in the main text (of volume $3.7\,\mu\mathrm{m}^3$ and $85\,\mu\mathrm{m}^3$, respectively). 
Note also that the edge path between the islands and the large cold electrodes is here 150$\,{\mu}\mathrm{m}$ long.

\textit{2) Experimental details.} 
The additional device is measured in the same dilution refrigerator.
It is immersed in a perpendicular magnetic field of $B\simeq2.22$\,T ($\nu=2$, center of plateau), $B\simeq4.46$\,T ($\nu=1$, center of plateau) and $B\simeq13.3$\,T ($\nu=1/3$, center of plateau).
The gains of the two noise amplification chains were obtained from thermal noise measurements, separately for each $B$.
In practice, the noise data on the two nominally identical islands are found indiscernible and are averaged together.
The on-chip electronic temperature $T_\mathrm{b}$ was determined from thermal noise.
Measurements at $B\simeq2.94$, 4.46 and 13.3\,T were performed, respectively, at $T_\mathrm{b}\simeq 12$, 12 and 21\,mK.

\textit{3) Stationary heat current.} 
The stationary heat current measured on the additional sample can be accounted for within experimental accuracy using the same electron-phonon power law $\alpha=5.5$ as for the sample in the main text, standard values of the electron-phonon prefactor $\Sigma=2.2$, 2.9 and 3.4\,$\times10^9$\,W.m$^{-3}$.K$^{-5.5}$ and an effective channel number for the thermal electronic heat flow of $N_\mathrm{eff}=3.2$, 1.4 and 1.8 for $B\simeq2.2$\,T ($\nu=2$), 4.5\,T ($\nu=1$) and 13.3\,T ($\nu=1/3$), respectively.

\textit{4) Heating dynamics.} 
Following the same procedure as in the main text, we first fit the measured relative amplitude of the slow thermalization stage (symbols in Fig.~\ref{fig-SI-Tstep-intermediate}(b)) with the prediction of Eq.~\eqref{Eq-RelativeStepHeight} (see also Eq.~\eqref{Eq-RelativeStepHeight-fitfn}) using $\gamma_\mathrm{ns}/K=0.085\pm0.009$\,J.s$^{-1}$.T$^{-2}$.m$^{-3}$ as a single adjustable parameter (lines of matching color).
Then we fit the observed characteristic time of the slow thermalization stage (symbols in Fig.~\ref{fig-SI-Tstep-intermediate}(a)) with the prediction of Eq.~\eqref{Eq-TauGeneral} (see also Eq.~\eqref{Eq-tau-fitfn}) using $K=0.21\pm0.03$ K.s as a single adjustable parameter.
The extracted nuclear spin parameters are found to be similar (30\% smaller) to those obtained from the main dataset on a different device.
This could come from a different amount of incorporated Ga, As or Al into the metallic island during the thermal annealing.
For a straightforward comparison, testing the predictive power of the simple model, we also display expectations with the same nuclear spin parameters as those used to account for the sample discussed in the main text (gray lines).
For a quantitative assessment, we can compare the variance characterizing the data-prediction comparison with the two sets of nuclear spin parameters. 
Using the nuclear spin parameters extracted from the first sample discussed in the main text, we find that the increase in the fit variance remains moderate with respect to the best fit of the present additional data: by a factor of 1.1 for $\tau$ and 2 for the slow stage relative amplitude.
Hence, the data-prediction agreement observed on this additional sample further confirms our observation and understanding of the two-step heating dynamics of mesoscopic islands and attests of the predictive power of the model.

\end{document}